\documentclass[10pt,a4paper,twoside]{article}
\usepackage{epsfig}
\usepackage{baltlat6}
\usepackage{array}
\usepackage{here}
\begin{document}
\ \ \vspace{0.5mm} \setcounter{page}{200}

\titlehead{Baltic Astronomy, vol.\,99, 999--999, 2014}

 \titleb{A NEW ESTIMATE OF THE LOCAL STANDARD OF REST FROM DATA ON YOUNG OBJECTS}

\begin{authorl}
\authorb{V.V. Bobylev}{1,2} and
\authorb{A.T. Bajkova}{1}
\end{authorl}

\begin{addressl}
 \addressb{1}{Central (Pulkovo) Astronomical Observatory of RAS, 65/1 Pulkovskoye Ch.,
 St. Petersburg, Russia; anisabajkova@rambler.ru}
 \addressb{2}{Sobolev Astronomical Institute,
 St. Petersburg State University, Bibliotechnaya pl.2,
 St. Petersburg, Russia; vbobylev@gao.spb.ru}
\end{addressl}

\submitb{Received: 2014 December 99; accepted: 2014 December 99}

\begin{summary}
 To estimate the peculiar velocity of the Sun with respect to the
Local Standard of Rest (LSR), we used young objects in the Solar
neighborhood with distance measurement errors within 10\%--15\%.
These objects were the nearest Hipparcos stars of spectral classes
O--B2.5, masers with trigonometric parallaxes measured by means of
VLBI, and two samples of the youngest and middle-aged Cepheids.
The most significant component of motion of all these stars is
induced by the spiral density wave. As a result of using all these
samples and taking into account the differential Galactic
rotation, as well as the influence of the spiral density wave, we
obtained the following components of the vector of the peculiar
velocity of the Sun with respect to the LSR:
 $(U_\odot,V_\odot,W_\odot)_\mathrm{LSR}=
 (6.0,10.6,6.5)\pm(0.5,0.8,0.3)\;\mathrm{km\,s}^{-1}$.
We have found that the Solar velocity components
 $(U_\odot)_\mathrm{LSR}$ and
 $(V_\odot)_\mathrm{LSR}$ are very sensitive to the Solar radial phase
$\chi_\odot$ in the spiral density wave.
\end{summary}

 \begin{keywords} Masers -- Galaxy: kinematics and dynamics -- galaxies: individual:
local standard of rest.
 \end{keywords}

%% \resthead is the RUNNING TITLE at top of the pages
 \resthead
 {A new estimate of the local standard of rest}
 {V.V. Bobylev, A.T. Bajkova}

\sectionb{1}{INTRODUCTION}
The peculiar velocity of the Sun with respect to the LSR
$(U_\odot,V_\odot,W_\odot)_\mathrm{LSR}$ plays an important role
in analysis of the kinematics of stars in the Galaxy. To properly
analyze Galactic orbits, this motion should be removed from the
observed velocities of stars, since it characterizes only the
Solar orbit~--- namely, its deviation from the purely circular
orbit. In particular, to build a Galactic orbit of the Sun, it is
desirable to know the components
$(U_\odot,V_\odot,W_\odot)_\mathrm{LSR}$.

There are several ways to determine the peculiar velocity of the
Sun with respect to the LSR. One of them is based on using the
Str$\ddot{o}$mberg relation. The method consists in finding such
values $(U_\odot,V_\odot,W_\odot)_\mathrm{LSR}$ that correspond to
zero stellar velocity dispersions
 (Dehnen \& Binney, 1998;
 Bobylev \& Bajkova, 2007;
 Co\c{c}kuno\u{g}lu et al., 2011;
 Golubov et al., 2013).
 This method was
addressed, for example, in the work by Sch\"onrich et al.~(2010),
where the gradient of metallicity of stars in the Galactic disk
was taken into account, and the velocity obtained is
 $(U_\odot,V_\odot,W_\odot)_\mathrm{LSR}=
 (11.1,12.2,7.3)\pm(0.7,0.5,0.4)\;\mathrm{km\,s}^{-1}$.

Another method implies a search for such
$(U_\odot,V_\odot,W_\odot)_\mathrm{LSR}$ that lead to minimal
stellar eccentricities. Using this approach,
 Francis \& Anderson (2009) suggested
that this velocity is
 $(U_\odot,V_\odot,W_\odot)_\mathrm{LSR}=
 (7.5,13.5,6.8)\pm(1.0,0.3,0.1)\;\mathrm{km\,s}^{-1}$.
To obtain this, 20000 local stars with known line-of-site
velocities were used. Using the improved database of proper
motions and line-of-site velocities of Hipparcos stars the same
authors found the following new values:
 $(U_\odot,V_\odot,W_\odot)_\mathrm{LSR}=
 (14.2,14.5,7.1)\pm(1.0,1.0,0.1)\;\mathrm{km\,s}^{-1}$
 (Francis \& Anderson, 2012);
 a great effort was made to get rid of the
influence of inhomogeneous distribution of velocities of stars
caused by kinematics of stellar groups and streams.

Another method is based on transferring stellar velocities towards
their origin. Following this approach, Koval' et al.~(2009)
derived the following values of
 $(U_\odot,V_\odot,W_\odot)_\mathrm{LSR}=
 (5.1,7.9,7.7)\pm(0.4,0.5,0.2)\;\mathrm{km\,s}^{-1}.$

Experience in using the Str$\ddot{o}$mberg relation showed that
the youngest stars significantly deviate from a linear dependence
when analyzing $(V_\odot)_\mathrm{LSR}$. This occurs when the
dispersions $\sigma<17$~km\,s$^{-1}$ (Dehnen \& Binney, 1998).
Therefore, the youngest stars are not normally used in this
method. Cepheids and other youngest objects fall into this area,
which allows us to include them in our analysis. This behavior of
velocities of the youngest stars is primarily connected with the
effect of the Galactic spiral density wave (Lin \& Shu, 1964). For
instance, the analysis of kinematics of 185 Galactic Cepheids by
Bobylev \& Bajkova~(2012) demonstrated that perturbation
velocities inferred by the spiral density wave can be determined
with high confidence.

The purpose of this paper is to estimate the velocity
$(U_\odot,V_\odot,W_\odot)_\mathrm{LSR}$ using spatial velocities
of the youngest objects in the Solar neighborhood that have
parallax errors not larger than 10\%--15\%. For these stars, we
consider not only the impact of the differential rotation of the
Galaxy, but also the influence of the spiral density wave.

\sectionb{2}{METHOD}\label{method}
Assuming that the angular rotation velocity of the Galaxy
($\Omega$) depends only on the distance  $R$ from the axis of
rotation, $\Omega=\Omega(R)$, the apparent velocity ${\bf V}(r)$
of a star at heliocentric radius ${\bf r}$ can be described in
vectorial notation by the following relation
 \begin{equation}
 {\bf V}(r)=-{\bf V}_\odot +
 {\bf V_\theta}(R)-{\bf V_\theta}(R_0) +
 {\bf V'},
 \label{Bottlinger-01}\end{equation}
where ${\bf V}_\odot(U_\odot,V_\odot,W_\odot)$ is the mean stellar
sample velocity due to the peculiar Solar motion with respect to
the LSR (hence its negative sign), the velocity $U$ is directed
towards the Galactic center, $V$ is in the direction of Galactic
rotation, $W$ is directed to the north Galactic pole; $R_0$ is the
Galactocentric distance of the Sun; $R$ is the distance of an
object from the Galactic rotation axis;
 ${\bf V_\theta}(R)$ is the circular velocity of the star with respect to the center
 of the Galaxy,
 ${\bf V_\theta}(R_0)$ is the circular velocity of the Sun, while
 ${\bf V'}$ are residual stellar velocities.

From the above relation~(\ref{Bottlinger-01}), one can write down
three equations in components ($V_r,V_l,V_b$), the so-called
\textit{Bottlinger's equations}~(Eq.~6.27 in Trumpler \& Weaver,
1953):
 \begin{equation}
 \begin{array}{lllll}
 V_r= (\Omega-\Omega_0)R_0\sin l \cos b,\\
 V_l= (\Omega-\Omega_0)R_0\cos l-\Omega r\cos b,\\
 V_b=-(\Omega-\Omega_0)R_0\sin l \sin b.
 \label{Bottl-0234}
 \end{array}
 \end{equation}
These are exact formulas, and the signs of $\Omega$ follow
Galactic rotation. After expanding $\Omega$ into Taylor series
against the small parameter $R-R_0$, then expanding the difference
$R-R_0$, where the distance $R$ is
 $R^2=r^2\cos^2 b-2R_0 r\cos b\cos l+R^2_0,$
and then substituting the result into Eq.~(\ref{Bottl-0234}), one
gets the equations of the Oort--Lindblad model (Eq.~6.34 in
Trumpler \& Weaver, 1953).

Our approach departs from the above in that the distances $r$ are
known quite well. In this case, there is no need to expand $R-R_0$
into series, since the distance $R$ is calculated using the
distances $r$. Furthermore, our approach implies an extra
assumption that the observed stellar velocities include
perturbations due to the spiral density wave ${\bf
V}_{sp}(V_R,\Delta V_\theta)$, with a linear dependence on both
${\bf V}_{sp}$ and ${\bf V}_\odot$. This allows us to write:
 $-{\bf V}_\odot=-{{\bf V}_\odot}_\mathrm{LSR}+{\bf V}_{sp}.$
Perturbations from the spiral density wave have a direct influence
on the peculiar Solar velocity ${{\bf V}_\odot}_\mathrm{LSR}$.
Then the relation~(\ref{Bottlinger-01}) takes the following form:
 \begin{equation}
 {\bf V}(r)=-{{\bf V}_\odot}_\mathrm{LSR}+{\bf V}_{sp} +
 {\bf V_\theta}(R)-{\bf V_\theta}(R_0) +
 {\bf V'},
 \label{Bottlinger-02}\end{equation}
which, considering the expansion of the angular velocity of
Galactic rotation $\Omega$ into series up to the second order of
$r/R_0$ reads
\begin{equation}
 \begin{array}{lll}
 V_r=-U_\odot\cos b\cos l-V_\odot\cos b\sin l-W_\odot\sin b\\
 +R_0(R-R_0)\sin l\cos b \Omega^\prime_0
 +0.5R_0 (R-R_0)^2 \sin l\cos b \Omega^{\prime\prime}_0\\
 +\Delta V_{\theta}\sin(l+\theta)\cos b-V_R \cos(l+\theta)\cos b,
 \label{EQ-1}
 \end{array}
 \end{equation}
 \begin{equation}
 \begin{array}{lll}
 V_l= U_\odot\sin l-V_\odot\cos l
  +(R-R_0)(R_0\cos l-r\cos b) \Omega^\prime_0\\
  +(R-R_0)^2 (R_0\cos l - r\cos b)0.5\Omega^{\prime\prime}_0- r \Omega_0 \cos b\\
  +\Delta V_{\theta} \cos(l+\theta)+V_R\sin(l+\theta),
 \label{EQ-2}
 \end{array}
 \end{equation}
 \begin{equation}
 \begin{array}{lll}
 V_b=U_\odot\cos l \sin b + V_\odot\sin l \sin b-W_\odot\cos b\\
    -R_0(R-R_0)\sin l\sin b\Omega^\prime_0
    -0.5R_0(R-R_0)^2\sin l\sin b\Omega^{\prime\prime}_0\\
    -\Delta V_{\theta} \sin (l+\theta)\sin b+V_R \cos(l+\theta)\sin b,
 \label{EQ-3}
 \end{array}
 \end{equation}
where the following designations are used: $V_r$ is the
line-of-sight velocity, $V_l=4.74 r \mu_l\cos b$ and $V_b=4.74 r
\mu_b$ are the proper motion velocity components in the $l$ and
$b$ directions, respectively, with the factor~4.74 being the
quotient of the number of kilometers in an astronomical unit and
the number of seconds in a tropical year; the star's proper motion
components $\mu_l\cos b$ and $\mu_b$ are in mas~yr$^{-1}$, and the
line-of-sight velocity $V_r$ is in km~s$^{-1}$; $\Omega_0$ is the
angular velocity at the distance $R_0$ from the rotation axis;
parameters $\Omega^\prime_0$ and $\Omega^{\prime\prime}_0$ are the
first and second derivatives of the angular velocity,
respectively. To account for the influence of the spiral density
wave, we used the simplest kinematic model based on the linear
density wave theory by Lin \& Shu~(1964), where the potential
perturbation is in the form of a travelling wave. Then,
 \begin{equation}
 \begin{array}{lll}
      V_R=f_R \cos \chi,\\
      \Delta V_{\theta}=f_\theta \sin \chi,
 \label{VR-Vtheta}
 \end{array}
 \end{equation}
where $f_R$ and $f_\theta$ are the amplitudes of the radial
(directed toward the Galactic center in the arm) and azimuthal
(directed along the Galactic rotation) velocity perturbations; $i$
is the spiral pitch angle ($i<0$ for winding spirals); $m$ is the
number of arms (we take $m=2$ in this paper); $\theta$ is the
star's position angle measured in the direction of Galactic
rotation: $\tan\theta=y/(R_0-x)$, where $x$ and $y$ are the
Galactic heliocentric rectangular coordinates of the object;
radial phase of the wave $\chi$ is
 \begin{equation}
   \chi=m[\cot (i)\ln (R/R_0)-\theta]+\chi_\odot,
 \label{chi-creze}
 \end{equation}
where $\chi_\odot$ is the radial phase of the Sun in the spiral
density wave; we measure this angle from the center of the
Carina--Sagittarius spiral arm ($R\approx7$~kpc). The parameter
$\lambda$, which is the distance along the Galactocentric radial
direction between adjacent segments of the spiral arms in the
Solar neighborhood (the wavelength of the spiral density wave), is
calculated from the relation
 $2\pi R_0/\lambda = m\cot(i).$
We take $R_0=8.0\pm0.4$~kpc, according to analysis of the most
recent determinations of this quantity in the review by Foster \&
Cooper~(20010).

In the present paper, we assume that parameters of both the
differential Galactic rotation and the spiral density wave are
known from observations of distant stars and solving equations of
the form~(\ref{EQ-1})--(\ref{EQ-3}). In this case the right-hand
parts of the equations contain only components of the Solar
peculiar velocity
\begin{equation}
 \begin{array}{lll}
 V_r-R_0(R-R_0)\sin l\cos b \Omega^\prime_0
 -0.5R_0 (R-R_0)^2 \sin l\cos b \Omega^{\prime\prime}_0\\
 -\Delta V_{\theta}\sin(l+\theta)\cos b-V_R \cos(l+\theta)\cos b\\
 =-U_\odot\cos b\cos l-V_\odot\cos b\sin l-W_\odot\sin b,
 \label{EQ-101}
 \end{array}
 \end{equation}
 \begin{equation}
 \begin{array}{lll}
 V_l-(R-R_0)(R_0\cos l-r\cos b) \Omega^\prime_0
    -(R-R_0)^2 (R_0\cos l - r\cos b)0.5\Omega^{\prime\prime}_0\\
 +r \Omega_0 \cos b-\Delta V_{\theta}\cos(l+\theta)-V_R\sin(l+\theta)
   =U_\odot\sin l-V_\odot\cos l,
 \label{EQ-102}
 \end{array}
 \end{equation}
 \begin{equation}
 \begin{array}{lll}
 V_b+R_0(R-R_0)\sin l\sin b\Omega^\prime_0
    +0.5R_0(R-R_0)^2\sin l\sin b\Omega^{\prime\prime}_0\\
    +\Delta V_{\theta} \sin (l+\theta)\sin b+V_R
    \cos(l+\theta)\sin b\\
  =U_\odot\cos l \sin b + V_\odot\sin l \sin b-W_\odot\cos b.
 \label{EQ-103}
 \end{array}
 \end{equation}
The system~(\ref{EQ-101})--(\ref{EQ-103}) can be solved by
least-squares adjustment with respect to three unknowns $U_\odot,$
$V_\odot$, and $W_\odot$. Another approach (which we follow) is to
calculate components of spatial velocities $U,V,W$ of stars:
 \begin{equation}
 \begin{array}{lll}
 U=V'_r\cos l\cos b-V'_l\sin l-V'_b\cos l\sin b,\\
 V=V'_r\sin l\cos b+V'_l\cos l-V'_b\sin l\sin b,\\
 W=V'_r\sin b                 +V'_b\cos b,
 \label{EQ-UVW}
 \end{array}
 \end{equation}
where $V'_r,V'_l,V'_b$ are left-hand parts of
Eqs.~(\ref{EQ-101})--(\ref{EQ-103}) which are the observed stellar
velocities free from Galactic rotation and the spiral density
wave. Then
 ${\overline U=-U_\odot},$
 ${\overline V=-V_\odot}$ and
 ${\overline W=-W_\odot}$.

\sectionb{3}{DATA}\label{Data}

\subsectionb{3.1}{O--B2.5 stars}\label{OB2stars}
The sample of selected 200 massive ($<$10$M_\odot$) stars of
spectral classes O--B2.5 is described in detail in our previous
paper (Bobylev \& Bajkova, 2013a). It contains spectral binary O
stars with reliable kinematic characteristics from the $3$~kpc
Solar neighborhood. In addition, the sample contains 124
Hipparcos~(van Leeuwen, 2007) stars of spectral types from B0 to
B2.5 whose parallaxes were determined to within 10\% and better
and for which there are line-of-sight velocities in the catalog by
Gontcharov~(20006).

In this work we solve the problem of determining the peculiar
velocity of the Sun. This problem can be solved most reliable
using the closest stars to the Sun. Therefore, from the database,
including 200 stars, we have selected 161 stars from the Solar
neighborhood of $0.7$~kpc radius.

Parameters of the Galactic rotation and the spiral density wave
used for reduction of motion of these stars were determined
in~(Bobylev \& Bajkova, 2013a) using full sample of 200 stars,
because for determination of Galactic parameters it is important
to have distant stars as well:
 $\Omega_0 = 32.4 \pm1.1$~km s$^{-1}$ kpc$^{-1},$
 $\Omega^\prime_0 = -4.33\pm0.19$~km s$^{-1}$ kpc$^{-2},$
 $\Omega^{\prime\prime}_0 = 0.77\pm0.42$~km s$^{-1}$ kpc$^{-3},$
 $f_R=-10.8\pm1.2$~km s$^{-1},$
 $f_\theta= 7.9\pm1.3$~km s$^{-1},$
 $\chi_\odot=-120^\circ\pm4^\circ.$
For all samples in the present work, we use the same value of the
wavelength $\lambda=2.6\pm0.2$~kpc ($i=-6.0\pm0.4^\circ$ for
$m=2$).

%%%%%%%%%%%%%%%%%%%%%%%%%%%%%%  FIGURE 1
\begin{figure}[!tH]
  \label{f1}
\vbox{
\centerline{\psfig{figure=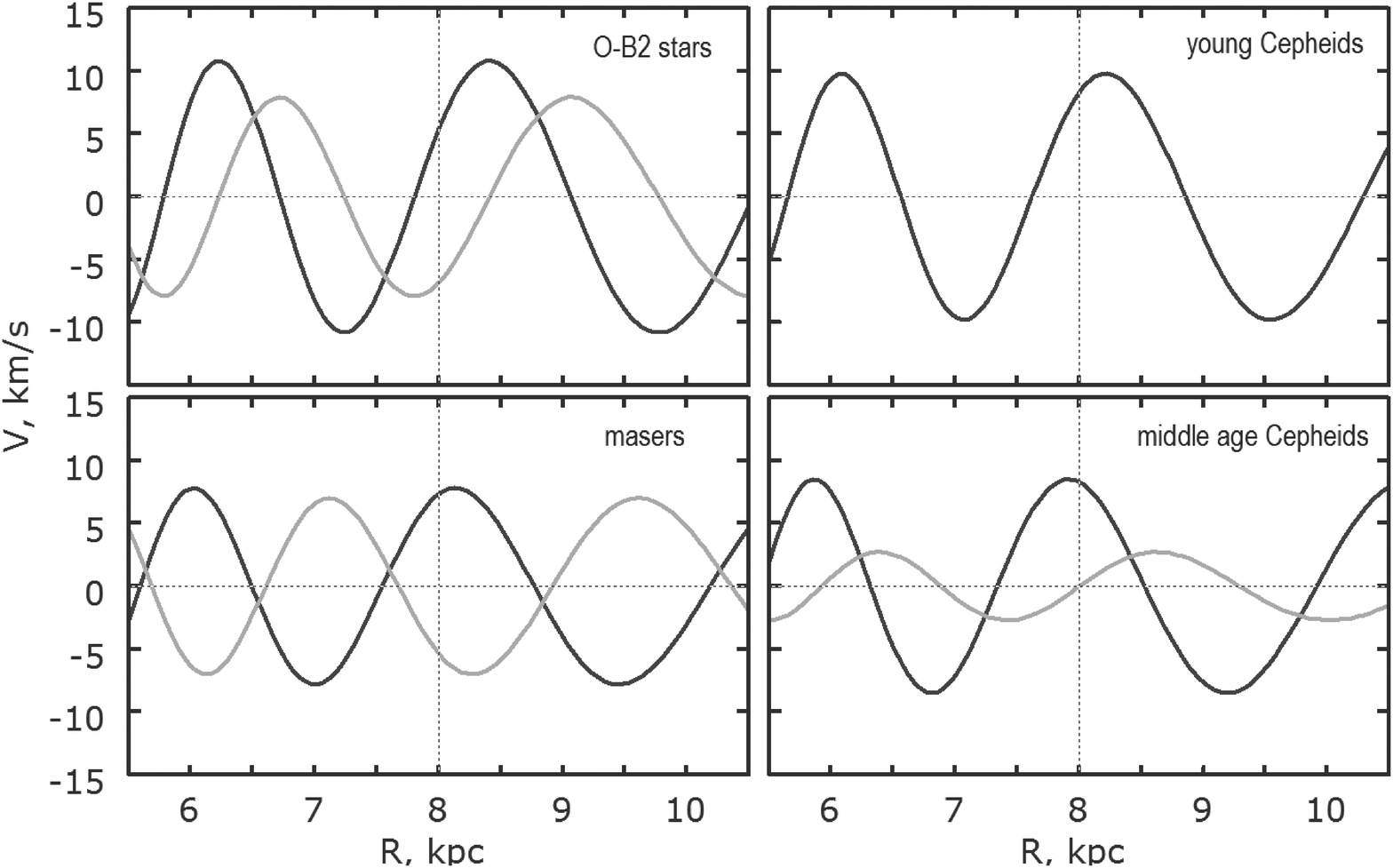,width=115mm,angle=0,clip=}}
\vspace{1mm} \captionb{1} {Radial ($V_R$, dark) and tangential
($\Delta V_\theta$, light)
 perturbation velocities versus Galactocentric distances $R$.
 Location of the Sun is indicated by a dotted line.} }
\end{figure}
%%%%%%%%%%%%%%%%%%%%%%%%%%%%%%%%%%%%%%%%%%

\subsectionb{3.2}{Masers}\label{masers}
We use coordinates and trigonometric parallaxes of masers measured
by VLBI with errors of less than 10\% in average. These masers are
connected with very young objects (basically proto stars of high
masses, but there are ones with low masses too; a number of
massive super giants are known as well) located in active
star-forming regions.

One of such observational campaigns is the Japanese project VERA
(VLBI Exploration of Radio Astrometry) for observations of water
(H$_2$O) Galactic masers at 22~GHz~(Hirota et al., 2007) and SiO
masers (which occur very rarely among young objects) at
43~GHz~(Kim et al., 2008). Water and methanol (CH$_3$OH) maser
parallaxes are observed in USA (VLBA) at 22~GHz and 12~GHz~(Reid
et al., 2009). Methanol masers are observed also in the framework
of the European VLBI network~(Rygl et al., 2010). Both these
projects are joined together in the BeSSeL program~(Brunthaler et
al., 2011). VLBI observations of radio stars in continuum at
8.4~GHz~(Dzib et al., 2011) are carried out with the same goals.

In the present work, we use only data on the nearby maser sources,
which are located no farther than 1.5~kpc from the Sun. All
required information about 30 such masers is given in the work
by~(Xu et al., 2013), which is dedicated to the study of the Local
arm (the Orion arm).

By applying the reduction algorithm, we use the following
parameters of the Galactic rotation and the spiral density wave
found by~(Bobylev \& Bajkova, 2013b):
 $\Omega_0 = 29.9\pm1.1$~km s$^{-1}$ kpc$^{-1},$
 $\Omega^\prime_0 = -4.27\pm0.20$~km s$^{-1}$ kpc$^{-2},$
 $\Omega^{\prime\prime}_0 = 0.915\pm0.166$~km s$^{-1}$ kpc$^{-3},$
 $f_R=-7.8\pm0.7$~km s$^{-1},$
 $f_\theta= 7.0\pm1.2$~km s$^{-1}$. In this case, the
values of the phase of the Sun in the spiral wave found
independently from radial and tangential perturbations using
Fourier analysis are different:
$(\chi_\odot)_R=-160^\circ\pm15^\circ$ and
$(\chi_\odot)_\theta=-50^\circ\pm15^\circ$, respectively.

Note that line-of-site velocities of masers given in the
literature usually refer to the standard apex of the Sun. So we
fix such line-of-site velocities, making them heliocentric.

\subsectionb{3.3}{Cepheids}\label{cepheids}
We used the data on classical Cepheids with proper motions mainly
from the Hipparcos catalog and line-of-sight velocities from the
various sources. The data from Mishurov et al.~(1997) and
Gontcharov (2006), as well as from the SIMBAD database, served as
the main sources of line-of-sight velocities for the Cepheids. For
several long-period Cepheids, we used their proper motions from
the TRC~(Hog et al., 1998) and UCAC4~(Zacharias et al., 2013)
catalogs. To calculate the Cepheid distances, we use the
calibration from Fouqu et al.~(2007), $\langle
M_V\rangle=-1.275-2.678\log P,$ where the period $P$ is in days.
Given $\langle M_V\rangle$, taking the period-averaged apparent
magnitudes $\langle V\rangle$ and extinction $A_V=3.23 E(\langle
B\rangle-\langle V\rangle)$ mainly from Acharova et al.~(2012)
and, for several stars, from Feast \& Whitelock~(1997), we
determine the distance $r$ from the relation
 \begin{equation}\displaystyle
 r=10^{\displaystyle -0.2(\langle M_V\rangle-\langle V\rangle-5+A_V)}
 \label{Ceph-02}
 \end{equation}
and then assume that the relative error of Cepheid distances
determined by this method is 10\%. We divided the entire sample
into two parts, depending on the pulsation period, which well
reflects the mean Cepheid age ($t$). We use the calibration from
Efremov~(2003),
 $\log t=8.50-0.65\log P,$
obtained by analyzing Cepheids in the Large Magellanic Cloud.
Parameters of the Galactic rotation and spiral density wave depend
on the age of the Cepheids. Therefore, for each sample of Cepheids
of the given age, these effects should be addressed individually.
We use the values of the parameters found in the work by Bobylev
\& Bajkova~(2012) for three age groups. The youngest Cepheids with
periods of $P\geq9^\mathrm{d}$ are characterized by the average
age of $55$~Myr, middle-aged Cepheids with periods of
$5^\mathrm{d}\leq P<9^\mathrm{d}$ have the average age of
$95$~Myr, while the oldest Cepheids with periods of
$P<5^\mathrm{d}$ have that of $135$~Myr. In the present work, a
sample of old Cepheids is not used because there are very few of
them in the Solar neighborhood, and their kinematic parameters are
not very reliable.

According to Bobylev \& Bajkova~(2012), for the youngest Cepheids
with periods of $P\geq9^\mathrm{d}$ $\Omega_0 = 26.1\pm0.9$~km
s$^{-1}$ kpc$^{-1}$,
 $\Omega^\prime_0 = -3.95\pm0.13$~km s$^{-1}$ kpc$^{-2}$,
 $\Omega^{\prime\prime}_0 = 0.79\pm0.10$~km s$^{-1}$ kpc$^{-3}$,
 $f_R=-9.8\pm1.3$~km s$^{-1}$,
 $\chi_\odot=-148^\circ\pm14^\circ$,
and the value of velocity perturbations in the tangential
direction $f_\theta$ is assumed to be zero. For middle-aged
Cepheids with $(5^\mathrm{d}\leq P<9^\mathrm{d})$ $\Omega_0 =
30.4\pm1.0$~km s$^{-1}$ kpc$^{-1}$,
 $\Omega^\prime_0 = -4.34\pm0.13$~km s$^{-1}$ kpc$^{-2}$,
 $\Omega^{\prime\prime}_0 = 0.69\pm0.14$~km s$^{-1}$ kpc$^{-3}$,
 $f_R=-8.5\pm1.1$~km s$^{-1}$,
 $f_\theta= 2.7\pm1.1$~km s$^{-1}$. The values for the phase of the
Sun in the spiral wave found separately from radial and tangential
perturbations by periodogram analysis based on Fourier transform
slightly differ: $(\chi_\odot)_R=-193^\circ\pm9^\circ$ and
$(\chi_\odot)_\theta=-180^\circ\pm9^\circ.$

%%%%%%%%%%%%%%
\begin{table}[!t]
\begin{center}
\vbox{\footnotesize\tabcolsep=3pt
\parbox[c]{124mm}{\baselineskip=10pt
{\smallbf\ \ Table 1.}{\small\ Components of the peculiar velocity
of the Sun with respect to the LSR, calculated considering the
differential Galactic rotation only.\lstrut}}
    \begin{tabular}{|l|c|c|c|c|l|c|c|c|}      \hline
                    Stars  &    $U_\odot$ &    $V_\odot$ &   $W_\odot$ & $N_\star$ & distance  \\
                           &  km s$^{-1}$ &  km s$^{-1}$ & km s$^{-1}$ &           &  kpc      \\\hline
 O--B2.5                   & $10.0\pm1.0$ & $14.7\pm1.3$ & $7.2\pm0.7$ &       161 & $<0.7$    \\
 masers                    & $11.9\pm2.7$ & $16.2\pm3.4$ & $6.2\pm1.7$ &        26 & $<1.5$    \\
 Cepheids,     $P\geq9^\mathrm{d}$  & $ 6.5\pm2.3$ & $12.0\pm2.4$ & $6.1\pm2.5$ &   36 & $<2$    \\
 Cepheids, $5^\mathrm{d}\leq P<9^\mathrm{d}$ & $ 7.3\pm2.1$ & $11.1\pm2.0$ & $6.4\pm1.8$ &   74 & $<2$ \\\hline
      \end{tabular}}
   \label{t1}
    \end{center}
   \end{table}
%%%%%%%%%%%%%%%%%%%%%%%%%%%%%%%%%%%%%

%%%%%%%%%%%%%%
\begin{table}[!t]
\begin{center}
\vbox{\footnotesize\tabcolsep=3pt
\parbox[c]{124mm}{\baselineskip=10pt
{\smallbf\ \ Table 2.}{\small\ Components of the vector of the
peculiar velocity of the Sun with respect to the LSR, calculated
considering both the differential Galactic rotation and the spiral
density wave.\lstrut}}
   \begin{tabular}{|l|c|c|c|c|l|c|c|c|}      \hline
                    Stars  &     $U_\odot$ &      $V_\odot$ &     $W_\odot$ & $N_\star$ & distance \\
                           &   km s$^{-1}$ &    km s$^{-1}$ &   km s$^{-1}$ &           &  kpc     \\\hline
 O--B2.5                   &   $4.6\pm0.7$ &   $ 8.6\pm0.9$ &   $7.2\pm0.7$ &       161 & $<0.7$   \\
 masers                    &   $6.0\pm1.6$ &   $11.4\pm2.5$ &   $6.2\pm1.7$ &        26 & $<1.5$   \\
 Cepheids,     $P\geq9^\mathrm{d}$  &   $6.8\pm2.3$ &   $12.1\pm2.4$ &   $6.1\pm2.5$ &        36 & $<2$  \\
 Cepheids, $5^\mathrm{d}\leq P<9^\mathrm{d}$ &   $6.7\pm2.1$ &   $10.4\pm1.9$ &   $6.4\pm1.8$ &        74 & $<2$  \\
                   average & $6.0\pm0.5$ & $10.6\pm0.8$ & $6.5\pm0.3$ &           &          \\\hline
      \end{tabular}
  \label{t2}
      }
    \end{center}
   \end{table}
%%%%%%%%%%%%%%%%%%%%%%%%%%%%%%%%%%%%%

 \sectionb{4}{RESULT AND DISCUSSION}\label{Results}

 \subsectionb{4.1}{The fixed value of $R_0$}\label{Approach1}
Here we describe the results obtained at fixed value of $R_0=8$
kpc, assuming the parameters of differential Galactic rotation and
the spiral density wave calculated earlier independently for each
stellar sample.

In Figure~1, there are radial ($V_R$) and tangential ($\Delta
V_\theta$) velocities of perturbations vs Galactocentric distance
$R$, induced by the spiral density wave. These velocities are
calculated according to the formulas (\ref{VR-Vtheta}) and
(\ref{chi-creze}) assuming $\theta=0^\circ$, and the amplitudes of
perturbations $f_R$ and $f_\theta$ defined in the data description
(Section~3). As it can be seen from this figure, at $R=R_0$, the
perturbations achieve about $5$~km~s$^{-1}$ in the radial
direction. In the tangential direction, the same value is achieved
for two samples: of youngest O--B2 stars and of masers. In the
case of young Cepheids, perturbations in the tangential direction
are not significant. In the case of middle-aged Cepheids,
perturbations in the tangential direction at $R=R_0$ are close to
zero. Note that a very small Solar neighborhood ($R\rightarrow
R_0$) is crucial to determine the velocity
$(U_\odot,V_\odot,W_\odot)_\mathrm{LSR}$.

In Table~1, the components of the peculiar velocity of the Sun
with respect to the LSR $(U_\odot,V_\odot,W_\odot)_\mathrm{LSR}$
are given. They were obtained only taking into account the
influence of the differential Galactic rotation. Components of
this vector, given in Table~2, were calculated considering both
the effects of the differential Galactic rotation and of the
spiral density wave.

As it is seen from Tables~1 and 2, considering the effect of the
spiral density wave for O--B2.5 stars and for masers leads to a
considerable variation of the components $\Delta U_\odot$ and
$\Delta V_\odot$ by $\approx$6~km$^{-1}$. In addition, this gives
smaller errors of the velocity
$(U_\odot,V_\odot,W_\odot)_\mathrm{LSR}$, which is especially
noticeable for masers.

The velocity $(V_\odot)_\mathrm{LSR}$ (Table~1) found from the
data on masers differs from
$(V_\odot)_\mathrm{LSR}=12.2\;\mathrm{km\,s}^{-1}$ (Sch\"onrich et
al., 2010) by $\approx$4~km\,s$^{-1}$, which is in accordance with
the result of analysis of masers in the Local arm (Xu et al.,
2013).

The following average values of the parameters
$(U_\odot,V_\odot,W_\odot)_\mathrm{LSR}$ found in the present work
are, essentially, more accurate than the estimate
 $(U_\odot,V_\odot,W_\odot)_\mathrm{LSR}=
 (5.5,11.0,8.5)\pm(2.2,1.7,1.2)\;\mathrm{km\,s}^{-1}$
obtained from 28 masers by Bobylev \& Bajkova~(2010) considering
the influence of the spiral density wave. The average value of
$(V_\odot)_\mathrm{LSR}$ (Table~2) is in a good agreement with the
result by Sch\"onrich et al.~(2010). There is a discrepancy in the
$(U_\odot)_\mathrm{LSR}$ component with Sch\"onrich et al.~(2010),
and especially with Francis \& Anderson~(2012).

Note that the revised Str$\ddot{o}$mberg relation applied to the
experimental RAVE data gives an absolutely different velocity
$(V_\odot)_\mathrm{LSR}\approx3\;\mathrm{km\,s}^{-1}$ (Golubov et
al., 2013). Using another approach to analysis of RAVE data
Pasetto et al.~(2012) obtained the following velocities:
$(U_\odot,V_\odot)_\mathrm{LSR}=
 (9.87,8.01)\pm(0.37,0.29)\;\mathrm{km\,s}^{-1}$.

Recently kinematic analysis of RAVE and the GCS (Nordstr\"om et
al. 2004) surveys was made by Sharma et al.~(2014). To constrain
kinematic parameters, were used analytic kinematic models based on
the Gaussian and Shu distribution functions.
 Sharma et al.~(2014) obtained the following velocities:
 $(U_\odot,V_\odot,W_\odot)_\mathrm{LSR}$ =
 $(10.96^{+0.14}_{-0.13}, 7.53^{+0.16}_{-0.16}, 7.539^{+0.095}_{-0.090})\;\mathrm{km\,s}^{-1}$.

Thus different methods give different results, and a final
agreement on the values of the velocity
$(U_\odot,V_\odot,W_\odot)_\mathrm{LSR}$ is not achieved till now.
We consider our estimates most reliable as they are based on the
youngest stars characterized by a small velocity dispersion and by
small Galactic orbit eccentricities as well.

%%%%%%%%%%%%%%
\begin{table}[!t]
\begin{center}
\vbox{\footnotesize\tabcolsep=3pt
\parbox[c]{124mm}{\baselineskip=10pt
{\smallbf\ \ Table 3.}{\small\ Components of the vector of the
peculiar velocity of the Sun with respect to the LSR, calculated
considering both the differential Galactic rotation and the spiral
density wave for the different values of $\chi_\odot$.\lstrut}}
 \label{t3}
   \begin{tabular}{|l|c|c|c|c|c|c|c|c|c|c|c|c|c|c|}      \hline
 $\chi_\odot$ & \multicolumn{2}{c|} {$-110^\circ$}&\multicolumn{2}{c|} {$-120^\circ$}&\multicolumn{2}{c|} {$-130^\circ$} \\\hline
              &             $U_\odot$&    $V_\odot$&            $U_\odot$&    $V_\odot$&            $U_\odot$&    $V_\odot$ \\
              &           km s$^{-1}$&  km s$^{-1}$&          km s$^{-1}$&  km s$^{-1}$&          km s$^{-1}$&  km s$^{-1}$ \\\hline
 O--B2.5      &           $6.3\pm0.5$& $ 8.0\pm0.6$&          $4.8\pm0.5$& $ 8.6\pm0.5$&          $3.5\pm0.5$& $ 9.3\pm0.3$ \\\hline
      \end{tabular}}
     \end{center}
   \end{table}
%%%%%%%%%%%%%%%%%%%%%%%%%%%%%%%%%%%%%

 \subsectionb{4.2}{Errors of Galactic Rotation Parameters}
Here we describe the results obtained for three particular values
of $R_0=7.5,$ $8.0,$ and $8.5$~kpc using the corresponding
differential Galactic rotation parameters. That is, we now use one
and the same Galactic rotation curve to analyze each of the
stellar samples. Amplitudes of perturbation velocities of the
spiral density wave $f_R$ and $f_\theta$, as well as the values of
the Solar phase $\chi_\odot$ in the spiral wave, are chosen as
above in Section~4.1.

For this purpose, we took a sample of masers (55 masers,
$\sigma_\pi/\pi<10\%,$ $r<3.5$~kpc) from Bobylev \&
Bajkova~(2013b) and, taking three fixed values of $R_0$, found the
following parameters of the Galactic rotation curve:
\begin{equation}
 \begin{array}{lll}
                 \Omega_0= 30.0\pm1.1~\hbox {km s$^{-1}$ kpc$^{-1},$}\\
         \Omega^\prime_0=-4.61\pm0.21~\hbox {km s$^{-1}$ kpc$^{-2},$}\\
 \Omega^{\prime\prime}_0=1.081\pm0.180~\hbox {km s$^{-1}$ kpc$^{-3},$}
 \quad R_0= 7.5~\hbox {kpc,}
  \label{R0-7.5}
 \end{array}
\end{equation}
\begin{equation}
 \begin{array}{lll}
                \Omega_0= 29.9\pm1.1~\hbox {km s$^{-1}$ kpc$^{-1},$}\\
         \Omega^\prime_0=-4.27\pm0.20~\hbox {km s$^{-1}$ kpc$^{-2},$}\\
 \Omega^{\prime\prime}_0=0.915\pm0.166~\hbox {km s$^{-1}$ kpc$^{-3},$}
 \quad R_0= 8.0~\hbox {kpc,}
  \label{R0-8}
 \end{array}
\end{equation}
\begin{equation}
 \begin{array}{lll}
                \Omega_0= 29.8\pm1.1~\hbox {km s$^{-1}$ kpc$^{-1},$}\\
         \Omega^\prime_0=-3.98\pm0.18~\hbox {km s$^{-1}$ kpc$^{-2},$}\\
 \Omega^{\prime\prime}_0=0.783\pm0.154~\hbox {km s$^{-1}$ kpc$^{-3}.$}
 \quad R_0= 8.5~\hbox {kpc,}
  \label{R0-8.5}
 \end{array}
\end{equation}
Parameters~(\ref{R0-8}) are the same as those used previously for
the maser sample in Section~4.1. %%%\ref{Approach1}.
Using three rotation curves (\ref{R0-7.5})--(\ref{R0-8.5}) we have
no found considerable departure from the previous results
(Table~2).

 \subsectionb{4.3}{Errors of the Spiral Wave Parameters}
Here we describe our results for several model values of the Solar
phase $\chi_\odot$ in the spiral density wave for a sample of
O--B2.5 stars (161 stars, $r<0.7$~kpc). We used the Galactic
rotation curve parameters~(\ref{R0-8}). The results are reflected
in Table~3 whence one can see that the Solar velocity components
$U_\odot$ and $V_\odot$ are very sensitive to the above parameter
($W_\odot$ velocities are not shown in the Table as they are
practically not affected by the density wave).

It is easy to understand these results by analyzing the
corresponding panel of Fig.~1 and Table~1. For instance, for
$\chi_\odot=-160^\circ$, the radial perturbation curve ($V_R$) is
near its maximum, so the influence to the $U_\odot$ component is
most prominent ($U_\odot=0.8\pm0.6$~km~s$^{-1}$). On the contrary,
the tangential perturbation curve ($\Delta V_\theta$) is about
zero, so there is no effect on the $V_\odot$ component
($V_\odot=12.5\pm0.5$~km~s$^{-1}$). For $\chi_\odot=-80^\circ$,
the radial perturbation curve ($V_R$) is near zero, so there is no
effect on the $U_\odot$ component,
$U_\odot=11.4\pm0.5$~km~s$^{-1}$. The tangential perturbation
curve ($\Delta V_\theta$) is near to minimum, so there is no
significant effect on the $V_\odot$ component,
$V_\odot=7.6\pm0.6$~km~s$^{-1}$.

We must note that, in our previous paper~(Bobylev \& Bajkova,
2013a), the uncertainty of $R_0$ was not taken into account when
determining the Solar phase in the spiral density wave
$\chi_\odot=-120\pm4^\circ$. We have redone Monte Carlo simulation
and obtained the following results:
 \begin{enumerate}
 \item
 If we consider only the error $\sigma_{R_0}=0.4$~kpc, its effect on the
uncertainty of the Solar phase in the spiral density wave becomes
very small: $\sigma_{\chi_\odot}=0.2^\circ$. The explanation for
this is that when you change $R_0,$ the length of a wave stretches
like a rubber band, but the phase of the Sun in the spiral wave
practically does not change.
 \item
 If we consider the errors of all observed parameters of
stars~-- parallaxes, proper motions, line-of-site velocities~--
along with the uncertainty $\sigma_{R_0}$, then the Solar phase in
the spiral density wave becomes $\chi_\odot=-120\pm6^\circ$.
 \end{enumerate}

Based on the data from Table~3, we may conclude that, in the range
of phase values from $-110^\circ$ to $-130^\circ$ (which is even
above the $1\sigma$ level), the Solar velocities in question,
found from O--B2.5 stars, are in the $U_\odot=6-4$~km~s$^{-1}$ and
$V_\odot=8-9$~km~s$^{-1}$ range.

\sectionb{5}{CONCLUSIONS}
For evaluation of the peculiar velocity of the Sun with respect to
the Local Standard of Rest, we used young objects from the Solar
neighborhood with distance errors of not larger than 10\%--15\%.
These are the nearest Hipparcos stars of spectral classes O--B2.5,
masers with trigonometric parallaxes measured by means of VLBI,
and two samples of the youngest and middle-aged Cepheids. The
whole sample consists of 297~stars. A significant fraction of
motion of these stars is caused by the Galactic spiral density
wave, because the amplitudes of perturbations in radial ($f_R$)
and tangential ($f_\theta$) directions reach $\approx$10~km
s$^{-1}.$

For each sample of stars, the impact of differential Galactic
rotation and of the Galactic spiral density wave was taken into
account. It was shown that, for the youngest objects~-- namely,
stars of spectral classes O--B2.5 and masers~-- considering the
effect of the spiral density wave leads to a change in the values
of the components of the peculiar velocity of the Sun with respect
to the LSR $\Delta U_\odot$ and $\Delta V_\odot$ by $\approx$6~km
$^{-1}$. Cepheids are less sensitive to the influence of the
spiral density wave.

Average values of the peculiar velocity of the Sun with respect to
the LSR are calculated according to the results of analysis of
four samples of stars; they have the following values:
 $(U_\odot,V_\odot,W_\odot)_\mathrm{LSR}=
 (6.0,10.6,6.5)\pm(0.5,0.8,0.3)$~km~s$^{-1}.$

We have found that components of the Solar velocity are quite
insensitive to errors of the distance $R_0$ in a broad range of
its values, from $R_0=7.5$~kpc to $R_0=8.5$~kpc, that affect the
Galactic rotation curve parameters. In the same time, the Solar
velocity components are very sensitive to the Solar phase
$\chi_\odot$ in the spiral density wave.

\thanks{
This work was supported by the ``Nonstationary Phenomena in
Objects of the Universe'' Program P--21 of the Presidium of the
Russian Academy of Sciences. }

 \References
 \refb
 Acharova I.A., Mishurov Yu.N., Kovtyukh V.V. 2012, MNRAS 420, 1590

\refb
 Bobylev V.V., Bajkova A.T. 2007, Astron. Rep., 51, 372

\refb
 Bobylev V.V., Bajkova A.T. 2010, MNRAS 408, 1788 %% 28 masers

\refb
 Bobylev V.V., Bajkova A.T. 2012, Astron. Lett. 38, 638 %% Cepheids

\refb
 Bobylev V.V., Bajkova A.T. 2013a, Astron. Lett. 39, 532 %% O-B2 stars

\refb
 Bobylev V.V., Bajkova A.T. 2013b, Astron. Lett. 39,  899 %% 73==55 Masers

\refb
 Bobylev V.V. 2013, Astron. Lett. 39,  909 %% Cepheids

\refb
 Brunthaler A., Reid M.J., Menten K.M., et al. 2011, AN 332, 461

\refb
 Co\c{c}kuno\u{g}lu B., Ak S., Bilir S., et al. 2011, MNRAS, 412, 1237

\refb
 Dehnen W., Binney J.J. 1998, MNRAS 298, 387

\refb
 Dzib S., Loinard L., Rodriguez L.F., et al. 2011, ApJ 733, 71

\refb
 Efremov Yu.N. 2003, Astron. Rep. 47, 1000

\refb
 Feast M., Whitelock P. 1997, MNRAS 291, 683

\refb
 Foster T., Cooper B. 2010, ASPC, 438, 16

\refb
 Fouqu P., Arriagada P., Storm J., et al. 2007, A\&A, 476, 73

\refb
 Francis C., Anderson E. 2009, New Astronomy, 14, 615

\refb
 Francis C., Anderson E. 2012, MNRAS 422, 1283

\refb
 Golubov O., Just A., Bienaym\'e O., et al. 2013, A\&A, 557, 92

\refb
 Gontcharov G.A. 2006, Astron. Lett. 32, 759

\refb
 Hirota T., Bushimata T., Choi Y.K., et al. 2007, PASJ 59, 897

\refb
 H{\o}g E., Kuzmin A., Bastian U., et al. 1998, A\&A, 335, L65

\refb
 Kim M.K., Hirota T., Honma M., et al. 2008, PASJ 60, 991

\refb
 Koval' V.V., Marsakov V.A., Borkova T.V. 2009, Astron. Rep. 53, 1117

\refb
 Lin C.C., Shu F.H. 1964, ApJ. 140, 646

\refb
 Mishurov Yu.N., Zenina I.A., Dambis A.K., et al. 1997, A\&A, 323, 775

 \refb
 Nordstr\"om B., Mayor M., Andersen J., et al. 2004, A\&A, 418, 989

 \refb
 Pasetto S., Grebel E.K., Zwitter T., et al. 2012, A\&A, 547, A7

\refb
 Reid M.J., Menten K.M., Zheng X.W., et al. 2009, ApJ 700, 137

\refb
 Rygl K.L.J., Brunthaler A., Reid M.J., et al. 2010, A\&A 511, A2

\refb
 Sch\"{o}nrich R., Binney J.J., Dehnen W. 2010, MNRAS, 403, 1829.

\refb
 Sharma S., Bland-Hawthorn J., Binney J., et al. 2014, ApJ, 793, 51

\refb
 Trumpler R.J., Weaver H.F. 1953, {\it Statistical Astronomy}.
 Univ. of Calif., Berkely

\refb
 Xu Y., Li J.J., Reid M.J., et al. 2013, ApJ 769, 15

\refb
 Zacharias N., Finch C., Girard T., et al. 2013, AJ, 145, 44

\end{document}